# Mapping Local Green Hydrogen Cost-Potentials by a Multidisciplinary Approach

S. Ishmam[1,2], H. Heinrichs[1,*], C. Winkler[1,2], B. Bayat[3], A. Lahnaoui[1], S. Agbo[5], E. U. Pena Sanchez[1,2], D. Franzmann[1,2], N. Ojieabou[1], C. Koerner[1], Y. Micheal[1], B. Oloruntoba[3], C. Montzka[3], H. Vereecken[3], H. Hendricks Franssen[3], J. Brendt[5], S. Brauner[1], W. Kuckshinrichs[4], S. Venghaus[1,6], D. Kone[7], B. Korgo[7,10], K. Ogunjobi[7], V. Chiteculo[8], J. Olwoch[8], Z. Getenga[9], J. Linßen[1] and D. Stolten[1,2]

1 Forschungszentrum Jülich GmbH, Institute of Energy and Climate Research – Jülich Systems Analysis (IEK-3), 52425 Jülich, Germany
2 RWTH Aachen University, Chair for Fuel Cells, Faculty of Mechanical Engineering, 52062 Aachen, Germany
3 Institute of Bio- and Geosciences: Agrosphere (IBG-3), Forschungszentrum Jülich GmbH, 52425 Jülich, Germany
4 Forschungszentrum Jülich GmbH, Institute of Energy and Climate Research – System Analysis and Technology Evaluation (IEK-STE), 52425 Jülich, Germany
5 Forschungszentrum Jülich GmbH, Corporate Development, (UE), 52425 Jülich, Germany.
6 RWTH Aachen University, School of Business and Economics, 52062 Aachen, Germany
7 West African Science Service Centre on Climate Change and Adapted Land Use (WASCAL), Agostino Road, PMB CT 504, Accra, Ghana
8 Southern African Science Service Centre for Climate Change and Adaptive Land Management (SASSCAL), 28 Robert Mugabe Avenue, Windhoek, Namibia.
9 Machakos University, Machakos County, Kenya
10 University Joseph KI-ZERBO, Ouagadougou, Burkina-Faso

* Corresponding author: h.heinrichs@fz-juelich.de

## Abstract

For fast-tracking climate change response, green hydrogen is key for achieving greenhouse gas neutral energy systems. Especially Sub-Saharan Africa can benefit from it enabling an increased access to clean energy through utilizing its beneficial conditions for renewable energies. However, developing green hydrogen strategies for Sub-Saharan Africa requires highly detailed and consistent information ranging from technical, environmental, economic, and social dimensions, which is currently lacking in literature. Therefore, this paper provides a comprehensive novel approach embedding the required range of disciplines to analyze green hydrogen cost-potentials in Sub-Saharan Africa. This approach stretches from a dedicated land eligibility based on local preferences, a location specific renewable energy simulation, locally derived sustainable groundwater limitations under climate change, an optimization of local hydrogen energy systems, and a socio-economic indicator-based impact analysis. The capability of the approach is shown for case study regions in Sub-Saharan Africa highlighting the need for a unified, interdisciplinary approach.

## 1 Introduction

In the wake of the rising cases of extreme weather conditions around the globe and the huge socio-economic losses and impacts driven by climate change, interventions relying on the use of green hydrogen as a decarbonization energy carrier has become a popular topic. At regional but also national and international levels, strategies to guide the production and use of hydrogen have been widely reported (Meng et al., 2021; Noussan et al., 2021). China for example has launched a hydrogen strategy targeting the transportation sector with measures to promote fuel cell vehicles (Noussan et al., 2021). The EU hydrogen strategy, which was adopted in 2022 prioritizes green hydrogen, and is committed to building a 40 GW electrolyzer capacity by 2030 (European Commission, 2020). It has the following five policy focusses: investment support, support production and demand, creating a hydrogen market and infrastructure, research and development, and international cooperation. The National Clean Hydrogen Strategy and Roadmap of the United States of America (US DOE, 2023) published in 2023 lays out pathways of leveraging clean hydrogen for the decarbonization of the various sectors of the economy. It plans to produce 10, 20 and 50 million metric tons of clean hydrogen annually by 2030, 2040 and 2050 respectively, resulting in a 10% reduction in greenhouse gas emission (US DOE, 2023). The German National Hydrogen Strategy was renewed in 2023 by the Federal Ministry for Economic Affairs and Climate Action (BMWK, 2023) detailing steps to accelerate the roll out of technology for the sustainable production and utilization of hydrogen. For its domestic industries, up to 130 TWh of hydrogen will be required by 2030 for which a 10 GW generation capacity for green hydrogen must be established. In total, 50-70% of the hydrogen demand is expected to be imported from European and international sources in the form of green hydrogen.

In Africa, green hydrogen strategies are on the rise as well. The Economic Community of West African states, ECOWAS launched her regional green hydrogen policy and strategy framework which has been adopted by the member states. It lays out that green hydrogen will play a key role for the subregion in aiding transition to green economy as well as mitigating climate change. It sets a short-term target of producing 0.5 million tons of green hydrogen annually by 2030 and a long-term target of 10 million tons by 2050 (ECREEE, 2023). Kenya launched her green hydrogen strategy in 2023 with the ambition to install 100 MW electrolysis capacity by 2027 and potential to scale up to 250 MW by 2032. The focus is for local green fertilizer production and for export (Burgess, 2023). All of these strategies combined with the huge renewable energy resources is to provide a workable framework for the implementation of hydrogen projects in Africa. Beyond these, the ECOWAS strategy for example makes it clear that there is the need for win-win partnerships to support the development of green hydrogen in Africa. These partnerships must recognize and consider local contexts through stakeholders' engagement in technology deployment and capacity building in order to guarantee local value and support. In Southern Africa, South Africa has been in the forefront of hydrogen development with the launch of its hydrogen research, development and innovation strategy, the so-called Hydrogen South Africa, HySA in 2008 (Pollet et al., 2014). HySA aims to position South Africa to actively drive hydrogen and fuel cell production and utilization along the entire value chain. HySA seeks to harness South Africa's mineral deposits and promote its cost-effective and sustainable use in the hydrogen economy and renewable energy. The Hydrogen Society Roadmap for South Africa 2021 builds further on HySA with a 70 action points to further boost the development of hydrogen in South Africa (DSI, 2021). The actions focus on hydrogen production, storage and distribution, decarbonization of transport, creation of export market among others (DSI, 2021). Namibia is another country in Southern Africa making giant strides in advancing green hydrogen technology. In its strategy document released in 2022, Namibia hopes to generate up to 12 million tons of hydrogen annually by 2050 from its three hydrogen valleys: southern region of Kharas, the central region including Walvis Bay port and the capital Windhoek, and the northern region of Kunene (Ministry of Mines and Energy Namibia, 2022).

Sub-Saharan Africa is a very promising region for green hydrogen production given its huge renewable energy resources. Africa has over 31 PWh/a renewable energy capacity (Mentis et al., 2015) that has mainly remained untapped. This huge potential can play a key role in drastically increasing local energy access. Beyond a part of this can be harnessed and utilized for green hydrogen production to balance variable renewable energy feed-in and for export. This way, the region has a clear path towards a 100% green energy access but also for economic opportunities arising from a green hydrogen economy.

However, special care must be taken to ensure the sustainability of water supply for green hydrogen production. Considering that for producing one kilogram of hydrogen, a consumption of nine kilograms of water is required (Beswick et al., 2021), a substantial supply of water resources is essential. Especially in regions, which already face substantial water security challenges, like those in Sub-Saharan Africa (Nkiaka et al., 2021). In this context, groundwater has gained widespread recognition as the primary water source sustaining diverse communities across the African continent (Altchenko et al., 2011).This acknowledgement is based on its ubiquitous presence, consistent availability, substantial storage capacity, superior water quality, and resilience to variations in both annual and seasonal climates, setting it apart from other available options (Adelana & MacDonald, 2008; Calow et al., 2010; Döll & Fiedler, 2008; MacDonald et al., 2012). Additionally, according to the United Nations Environment Programme (UNEP) in 2010, the large groundwater storage capacity solidifies its status as the most abundant water resource in Africa (UNEP, 2010). These characteristics make groundwater an attractive and essential source for sustainable new energy production in Africa. A thorough assessment of groundwater availability in Africa is considered not only to foster beneficial hydrogen production, but also to implement sustainable water consumption strategies to prevent the overexploitation and depletion of groundwater reserves (Bierkens & Wada, 2019; Custodio, 2002; de Graaf et al., 2019; Rohde et al., 2018), and avoid potential conflicts among diverse water users. To unlock the option for green hydrogen production in regions without sustainable groundwater the option to supply water through seawater desalination has gained much attention as well (Franzmann et al., 2023).

In each of the touched research areas several works have been done, even if not jointly across the decisive dimensions for a green hydrogen economy in Sub-Saharan Africa. Starting with studies assessing renewable energy potentials in that region, a lot of work has been done to quantify the available renewable energy resources in Africa in general. Bastian Winkler et al (2017) implemented the so-called Integrated Renewable Energy Potential Assessment for assessing smallholder farming systems in South Africa. The authors opined that consideration should be given to the various inter-related social, environment and technological factors to be able to properly estimate RE potential. Spatially explicit models and long-term satellite data have also been used to assess offshore wind technical potentials (Elsner, 2019). Especially for the African continent, Mentis et al. (Mentis et al., 2015) estimate that an annual potential from onshore wind generation could reach up to 31 PWh based on the technical potential, but only calculates timeseries daily. Within a global assessment, Bosch et al. (Bosch et al., 2017) determines annual time series for Africa, combining simulation methods from weather data with a suitability approach for potential estimation. For PV, global studies for technical potentials exist as well like Pietzcker et al. (Pietzcker et al., 2014) and Köberle et al. (Köberle et al., 2015). For the ECOWAS region, Yushchenko et al. (Yushchenko et al., 2018) analyzed the potentials for open field PV using a multi criteria decision analysis combining technical potential evaluation with feasibility factors like distances to grid connections or roads and therefore not limiting the potentials to a technical feasibility.

For African countries, only partial studies have been conducted to determine the hydrogen generation. Franzmann et al. (Franzmann et al., 2023) show for single selected countries, that liquid hydrogen can be exported at quantities greater than 1 PWh/a at cost starting at 2 EUR/$kg_{H2}$ (Franzmann et al., 2023). They evaluate the hydrogen costs at a given export location, considering the cost of transportation. IRENA (IRENA, 2022) determines the local gaseous hydrogen cost for the African continent starting at 1.1 USD/$kg_{H2}$. In addition to hydrogen potentials, Mukelabai et al. (Mukelabai et al., 2022) show, that for producing green hydrogen from renewables, water limitations in Sub-Saharan-Africa play a crucial role for electrolysis.

Understanding how the renewable energy resources can be relevant for green hydrogen production as well as knowing how much of green hydrogen production is possible play a significant role in addressing political and socio-cultural challenges towards integrating hydrogen in the national energy mix of Sub-Saharan countries. An approach for such a study therefore requires detailed and in-depth assessment of the renewable energy resources considering the prevailing local contexts and preferences derived from extensive socio-economic, techno-economic, ecological and stakeholder engagement. The uniqueness of land use and tenure system in Africa makes it imperative that a well-articulated land eligibility assessment (LEA) for hydrogen production is carried out in close consultation with local stakeholders.

To ensure a sustainable water supply, our attention is directed towards the sustainable yield of groundwater. This refers to the amount of groundwater available for extraction that can be utilized over an extended period without causing negative effects, while also maximizing economic, social, and environmental benefits (Alley & Leake, 2004; Fetter, 1972, 2001; Freeze, 1971; Kalf & Woolley, 2005; Shi et al., 2012; Sophocleous, 2000; Sophocleous & Perkins, 2000). To quantify the groundwater sustainable yield, it is essential to consider groundwater recharge in addition to total human water usage and environmental flow. On a global scale, the initial groundwater recharge study was conducted in 1979 by L'vovič (L’vovič, 1979). This study utilized a baseflow component of measured river discharge to create a comprehensive global map of groundwater recharge (L'vovich, 1979). Subsequently, global groundwater recharge has primarily been estimated through the utilization of hydrological models. For instance, Döll et al. (Döll et al., 2002) generated a global groundwater recharge map by means of the hydrological model WGHM (WaterGAP Global Hydrology Model (Alcamo et al., 2003; Döll et al., 2003)), and updated simulations with the WGHM2 model (Döll & Fiedler, 2008). Considerable research efforts have been dedicated to investigating various facets of groundwater recharge across numerous regions in Africa, for instance, in southern Africa (Abiye, 2016; Xu & Beekman, 2003), northern Africa (Edmunds & Wright, 1979; Guendouz et al., 2003; Sturchio et al., 2004) and western Africa (Edmunds & Gaye, 1994; Favreau et al., 2009; Leblanc et al., 2008; Leduc et al., 2001). The first comprehensive long-term groundwater recharge map covering the entirety of Africa was produced, spanning from 1970 to 2019. This map was created using estimates gathered from ground-based measurements (MacDonald et al., 2021). More recently, five decades (1965-2014) of groundwater recharge simulations have been performed for African continent to investigate groundwater sustainability and encourage its sustainable consumption (Bayat et al., 2023).

Regarding the socioeconomic dimension renewable energy (RE) holds the potential for positive impacts (del Río & Burguillo, 2009). To accurately measure and assess these impacts, it is imperative to employ suitable indicators. Composite indicators (CIs) provide a valuable means of aggregating the numerous individual factors inherent in each case, thereby enhancing measurability and analyzability (Zhou et al., 2007). Recent research indicates a continual expansion and refinement in the application of composite indicators. For instance, a study by Borbonus (Borbonus, 2017) introduces a framework that dissects the potential social and economic benefits of renewables into subcategories that can be adapted to the specific needs of each country. Another framework presents a scalable and internationally transferable methodology for a composite index of the risk of energy poverty in domestic heating (Kelly et al., 2020). Further examples of the utility of CIs in addressing complex issues such as renewable energy include a study identifying connections between access to electricity and social development (Zhang et al., 2019). The authors conclude, based on their indicator analysis, that "electrification of social infrastructure can serve as an energy anchor, facilitating and enlarging electrification of neighborhood communities and productive use (Casati et al., 2023)". Research assessing an energy poverty index indicates that the broader implementation of sustainable energy sources can contribute to alleviating energy poverty (Zhao et al., 2022). Similar findings were also uncovered in a CI-based study examining the impact of RE production at specific levels of urbanization (Lantz et al., 2021).

The work reported here follows an extensive investigation into the potential of producing green hydrogen aiming especially at countries in sub-Saharan Africa. The aim is to present the detailed approach that has been employed in determining the total green hydrogen production potential and the unit cost exemplified in a region in Sub-Saharan Africa. It details the steps taken in ensuring that every aspect of technology, environment and local scenarios as described above are considered combining both theoretical, analytical, experimental, and social approaches, The underlying methodology is presented in Section 2 and the types of achieved results are introduced in Section 3. Section 4 critically reflects on and concludes about the applied approach and obtainable results.

## 2 Methodology

The underlying methodology of this study consists of several coupled models and integrated data to finally achieve its given goal to serve as decision support shown here for an exemplarily analyzed region in Sub-Saharan Africa. It starts with a dedicated land eligibility analysis which identifies where renewable energy technologies can be placed in accordance with local preferences. Based on those placements each location specific wind turbine and open-field photovoltaic module are simulated to derive their weather dependent hourly electricity generation. In parallel, the availability of sustainable groundwater and the cost for seawater desalination are calculated. In a next step, both water supply options together with the time series of renewable energy technologies are fed into an energy system model for each “GID-2” administrative region based on the definition from Global Administrative Areas (Global Administrative Areas, 2020) in the analyzed regions. Each energy system model additionally consists of batteries as potential electricity storage, hydropower plants and electrolysis to supply an exogenously increasing hydrogen demand until the maximum potential is achieved to form a green hydrogen cost-potential curve for each region. In addition to these techno-economic results, the methodology was extended by drawing on previous studies that overlooked aspects related primarily to socio-technical and socio-economic factors. Indeed, since energy projects would provide local added value and could be beneficial in those regions, where the energy supply is not yet fully developed, opportunities for the population should be explored. Together these results are made available via a web-based graphical user interface (GUI) to be accessible for a wide range of decision-makers and stakeholders.

### 2.1 Land eligibility assessment for open-field photovoltaic and onshore wind turbines

A land eligibility analysis that yields the overall land area eligible for the placement of renewable plants under specified constraints usually comes before determining the theoretical maximum renewable potential like for onshore wind (Deng et al., 2015; McKenna et al., 2015; Zappa & van den Broek, 2018). Typically, the land eligibility analysis takes into account various factors such as topography, land use, environmental constraints, and social and economic considerations (Ryberg et al., 2018). Within the context of this study, the set of land eligibility criteria have been carefully selected from literature sources (Franzmann et al., 2023; Ryberg et al., 2018) and complemented by feedback received from our local partners. As this study focuses specifically on Sub-Saharan Africa, the inclinations of regional stakeholders, such as community members, governmental bodies, and international institutions are of utmost importance to achieve meaningful results for decision makers. Therefore, deliberate efforts have been made to collect, process, and examine the local preferences aligned with the criteria listed in Table 1 within several workshops. The set of constraints thus obtained are referred to as "local preferences" in the subsequent sections. In cases where it was not possible to receive stakeholder preferences, the median of the local preferences collected was allocated. This has been done for onshore wind turbines and open-field photovoltaics, which are perceived as major building blocks of the global energy transition with comparable low environmental impact (International Renewable Energy Agency, 2023). This approach results in 33 criteria and buffer distances for onshore wind and open-field PV each.

The research was conducted using the open-source general-purpose geospatial toolkit GeoKit (GeoKit, 2024) and the land eligibility model “Geospatial Land Availability for Energy Systems” (GLAES) (GLAES, 2024). GLAES leverages the capabilities of GeoKit to perform all its underlying geospatial operations and is specifically designed for land eligibility analysis. Its design focuses on reducing errors that may arise from geospatial operations commonly needed for land eligibility analysis, being transparent in methodology, scalable for vast geographical regions, and adaptable to standard geospatial data formats. Both GeoKit and GLAES are implemented in the Python 3 programming language and rely on the SciPy (Virtanen et al., 2020) ecosystem for general numerical and matrix computations and the Geospatial Data Abstraction Library (GDAL) (GDAL/OGR contributors, 2024) for geospatial operations, both of which are open-source projects. The computations for the land eligibility analysis were conducted using a custom Lambert azimuthal equal-area projection (LAEA) projection system with a 100 m spatial resolution centered on each region individually, which provide the highest accuracy for the area determination. The exclusion features are the most comprehensive in comparable studies to date. In combination with carefully selected high-resolution geospatial datasets that are particularly suitable for the project region, the resulting land eligibility has an unprecedented degree of reliability. This allows to show geospatial exclusions even against high resolution satellite imagery. However, it is not able to capture exclusions that are not clearly defined by geospatial properties, such as the objection of landowners or lawsuits due to for example endangered species which are found in the area only upon local inspection.

Figure 1 displays an exemplary case of how land exclusions correspond to each criterion in a region in Sub-Saharan Africa. The application of each criterion, with its associated buffers, such as 200 m around primary roadways, 1000 m around settlements, etc., results in the progressive exclusion of more and more land areas. As a result, only a mere 23.8% of the land area is available for this exemplary region.

Based on the collected local preferences, land exclusions were applied in for all the 33 criteria and eligible areas were computed for each region for both wind turbines and photovoltaic modules. The resulting eligible areas form the basis for the renewable energy potential assessment in Section 2.2.

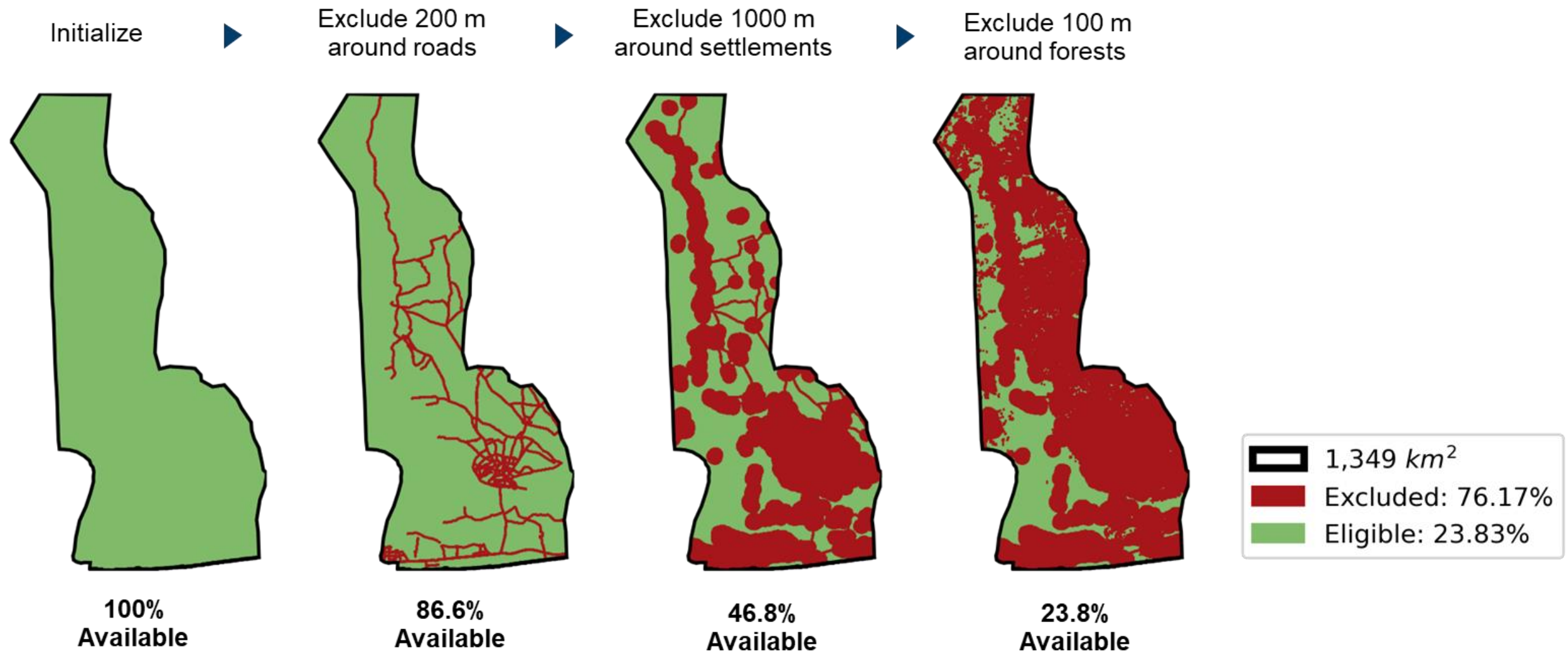

**Figure 1. Exemplary exclusion of different criteria for an example region in western Africa (Ouémé in Benin)**

**Table 1. Selected criteria for land eligibility analysis**

| Criteria | | | |
|---|---|---|---|
| 1 | Settlements (connected) | 18 | Lakes |
| 2 | Settlements (isolated) | 19 | Creeks |
| 3 | Airports | 20 | Rivers |
| 4 | Primary Roadways | 21 | Coastlines (Ocean, general) |
| 5 | Secondary Roadways | 22 | Woodlands (All Forests) |
| 6 | Agricultural Areas | 23 | (Standard) Wetlands |
| 7 | Pasture Areas | 24 | Specially protected Wetlands |
| 8 | Railways | 25 | Sand Dunes |
| 9 | Power Lines | 26 | Natural Habitats |
| 10 | Historical Sites | 27 | Biospheres |
| 11 | Recreational Areas | 28 | Wildernesses |
| 12 | Leisure and Camping | 29 | Bird Areas |
| 13 | Industrial Areas | 30 | Protected Landscapes |
| 14 | Commercial Areas | 31 | Natural Reserves |
| 15 | Mining Sites | 32 | National Parks, State Parks, etc. |
| 16 | Military Areas | 33 | Natural Monuments |
| 17 | National Borders | | |

## 2.2 Renewable energy potential assessment

Subsequently, wind turbines and open-field PV modules are placed on the eligible areas. For this wind turbines are positioned with a distance of 8 times the rotor diameter for the main wind speed direction and 4 times the rotor diameter for the transversal wind speed direction using the placement algorithm in GLAES (Ryberg et al., 2019). The turbines utilized here are specifically designed for future applications and are extensively examined in Ryberg et al. (Ryberg et al., 2019), as the applied scenario is specifically framed for the year 2050. The investment cost of each individual turbine is determined through the utilization of a cost model. Hourly simulations are then conducted using the appropriate meteorological data from the ERA5 weather data source (Hersbach et al., 2018) for the corresponding year, following the methods outlined in Ryberg et al. based on the open-source tool RESkit (RESKit, 2024; Ryberg et al., 2019).

Similarly, simulation sites for open-field PV parks are placed with a minimum distance of 1000 meters between them. When considering PV parks, the area surrounding each placement is an important factor in its potential capacity, as they are much more spread out than wind parks. Eligible area in the case of PV parks is assigned by creating Voronoi polygons (Okabe et al., 2009) at a size of 1000 m around each possible placement location and grouping adjacent and non-contiguous eligible area together. As a last step, the placement locations are updated to correspond to the centroid of each identified area. For this technical potential an open-field coverage of 20 $m^2/kW_p$ and the module Winaico WSx-240P6, as suggested by Ryberg (Ryberg et al., 2020) is assumed. Finally, all the locations are simulated for open-field PV systems without single-axis tracking, using the algorithm of RESkit described in (Ryberg et al., 2020). We selected a fixed PV system as it has the lowest investment cost, which is especially relevant for countries potentially facing financing challenges.

The hydropower potentials in Africa from Sterl et. al. (Sterl et al., 2021) were utilized to derive the electricity generation time-series available from existing and planned hydropower plants in Africa. The database offers monthly hydropower generation data under three distinct scenarios: "dry," "normal," and "wet" seasons. For this analysis, the "normal" scenario was adopted as it represents a standard performance of the hydropower hydro-fleet according to the author. In addition, its use allows for hydropower production estimation potential under no extreme conditions. Only hydropower plants with a capacity exceeding 1 MW were included in the analysis, encompassing both run-of-river and reservoir hydropower facilities. This makes 394 medium and large hydropower plants total with 84 GW installed capacity for 2050 considered in the assessment. To align with the temporal resolution of wind turbines and PV modules, the monthly generation time series was transformed into an hourly form by linearly projecting the monthly generation into hourly values from one month to the next. The monthly mean values were assigned in the middle of the month. Hydropower costs were taken from the report “Renewable Power Generation Costs in 2019” from IRENA (International Renewable Energy Agency, 2020).

The geothermal potentials are based on a methodology from Franzmann et al. (Franzmann et al., 2024). Based on the land eligibility results from section 2.1 for geothermal plants, the capacity for enhanced geothermal systems is calculated. The geological temperature is estimated based on an approach from Aghahosseini et al. (Aghahosseini & Breyer, 2020) and heat flow data from Goutorbe  et al. (Goutorbe et al., 2011). The heat output of the plants is calculated based on Gringartens approach (Augustine, 2017; Gringarten & Sauty, 1975) applied to location specific plants and the thermal conversion is based on Tester (Tester, 2006). As geothermal cost is mostly dependent on drilling costs, the medium outlook of geothermal drilling costs is assumed based on NREL (Mines, 2016) for a maximum depth of 7 km (Tester, 2006).

The resulting time series of electricity generation of each wind turbine, open-field PV park and hydropower plants form the supply basis for the subsequent step to derive the green hydrogen potential.

### 2.3 Sustainable water supply assessment

For the assessment of water availability, we considered sustainable groundwater supply and seawater desalination including water transport. By considering both options, all regions have a potential source of water supply.

#### 2.3.1 Sustainable groundwater supply

*Groundwater recharge*

The annual groundwater recharge was determined through simulations using the Community Land Model [ver. 5] (CLM5) (Lawrence et al., 2019) employing a general water balance method (Bayat et al., 2023; Hahn et al., 1997; Meinzer, 1920; Rossi et al., 2022). Furthermore, to account for anthropogenic water inputs, the water balance integrated irrigation, following CLM simulations (Ozdogan et al., 2010), as indicated by:

$$R = (P + I) - ET - Q \tag{1}$$

where $R$ is groundwater recharge [mm yr$^{-1}$], $P$ is the precipitation (rain and snow) [mm yr$^{-1}$], $I$ is the simulated irrigation by CLM to account for all anthropocentric water supply [mm yr$^{-1}$], $ET$ is evapotranspiration [mm yr$^{-1}$], and $Q$ is surface runoff [mm yr$^{-1}$]. Groundwater recharge was estimated on a yearly basis.

*Environmental flow*

Environmental flow (Qrest), or the minimum ecological water requirement, represents the essential water volume crucial for sustaining ecosystems and the services they provide. Previous recommendations advocate that the responsible use of water resources should not exceed 10%, 40%, and 70% of the total groundwater recharge, aligning with conservative, medium, and extreme scenarios (Alley et al., 1999; Maimone, 2004; Shi et al., 2012; Sophocleous, 2000). In accordance with these guidelines, we have adopted three distinct scenarios to allocate environmental flow based on the simulated groundwater recharge: (i) a conservative scenario (90% of recharge), (ii) a medium scenario (60% of recharge), and (iii) an extreme scenario (less conservative with 30% of recharge).

*Sustainability analysis*

Determining the groundwater sustainable yield involves the application of Eq. (2), which incorporates the concept of the percentage of recharge, as suggested by (Hahn et al., 1997; Miles & Chambet, 1995), alongside the integration of total sectoral water consumption. This method enables the assessment of the sustainability of groundwater utilization.

$$SY = R - Qrest - SWU \tag{2}$$

where SY is groundwater sustainable yield, $Q_{rest}$ is environmental flow, and SWU is sectoral water use. The units of all variables in Eq. (2) can be expressed in mm $yr^{-1}$.

Moreover, we have considered three distinct scenarios for the sustainable yield derived from the three environmental flow scenarios (conservative, medium, and extreme). For groundwater availability analysis, the long-term (2015 - 2100) groundwater sustainable yield has been calculated, and then the average from 2015 – 2035, 2015 – 2045 and 2036 – 2065 are considered representative of 2020, 2030 and 2050. The groundwater recharge is obtained from the CLM5 model (Lawrence et al., 2019) simulations at 10 km spatial resolution, forced by the regional climate model runs of CCLM5 (Sørland et al., 2021), REGCM4 (Coppola et al., 2021), and REMO2015 (Pietikäinen et al., 2018), which were driven by the MPI (Gutjahr et al., 2019) and NOR-ESM (Bentsen et al., 2013) General Circulation Models (GCMs). The RCP2.6 (which is considered optimistic indicating low greenhouse gas concentration pathways) and RCP8.5 (which is considered pessimistic indicating high greenhouse gas concentration pathways) scenarios were considered for these simulations. The average groundwater recharge (from six combinations of GCMs and RCMs), and consequently the groundwater sustainable yield, is considered for further use in the green hydrogen project for RCP2.6 (optimistic) and RCP8.5 (pessimistic) scenarios. The environmental flow is calculated from simulated groundwater recharge. The sectoral water use is obtained from literature (Sutanudjaja et al., 2018) that takes total water withdrawal for industrial, irrigation, domestic, and livestock into account.

### 2.3.2 Desalinating seawater and water transport

The selection of groundwater or desalinated water for hydrogen production depends on the cost incurred. Since groundwater is usually the cheaper option, this means that local groundwater resources are considered first, and desalinated seawater is used once the limits of sustainable groundwater are reached. Regions that do not have sustainable groundwater resources left have to use exclusively desalinated water. After a comparison of published literature on water transport cost, a validated UN cost model (United Nations ESCWA, 2009) based on real data (Kelly et al., 2020) was selected to integrate water supply from seawater desalination in our analysis. This updated model requires the electricity cost for pumping as well as elevation and distance to the shore as an input. This base model was further developed by updating cost values and converting diesel- to electricity-powered pumps. Electricity cost was introduced as a variable and set to the average solar photovoltaic levelized cost for 24-hour operation: Exemplary studies in the affected regions revealed that, under the given techno-economic assumptions, the unit cost for constant operation amounts on average to ca. 3x the solar electricity production cost due to the required extra cost for battery storage. Transport distance is then calculated based on distance of the district centroid from the nearest coastline (GADM, 2023) and a detour factor of 1.3 typical for infrastructure routing (Reuß et al., 2021). Elevation is extracted as the average regional value from OpenTopography (OpenTopography, 2016). Desalinated water cost is calculated using the cost model of Loutatidou et al. (Loutatidou et al., 2014), scaled to present day cost, with the previously mentioned PV-based electricity cost and a reference plant size of 367 000 m³/h, based on Heinrichs et al. (Heinrichs et al., 2021).

## 2.4 Local green hydrogen potential assessment

In this study the technical hydrogen potential for each "GID_2" region has been evaluated. The "GID_2" regions refer to the administrative divisions of the second level as categorized by the GADM (Database of Global Administrative Areas) (Global Administrative Areas, 2020). In countries where no “GID_2” level is available, the first level “GID_1” divisions were used. In the following analysis, the simplified term “GID_2 regions” will be used for all regions for improved readability. For each of these GID-2 regions a separate energy system model was defined containing the components shown in Figure 2.

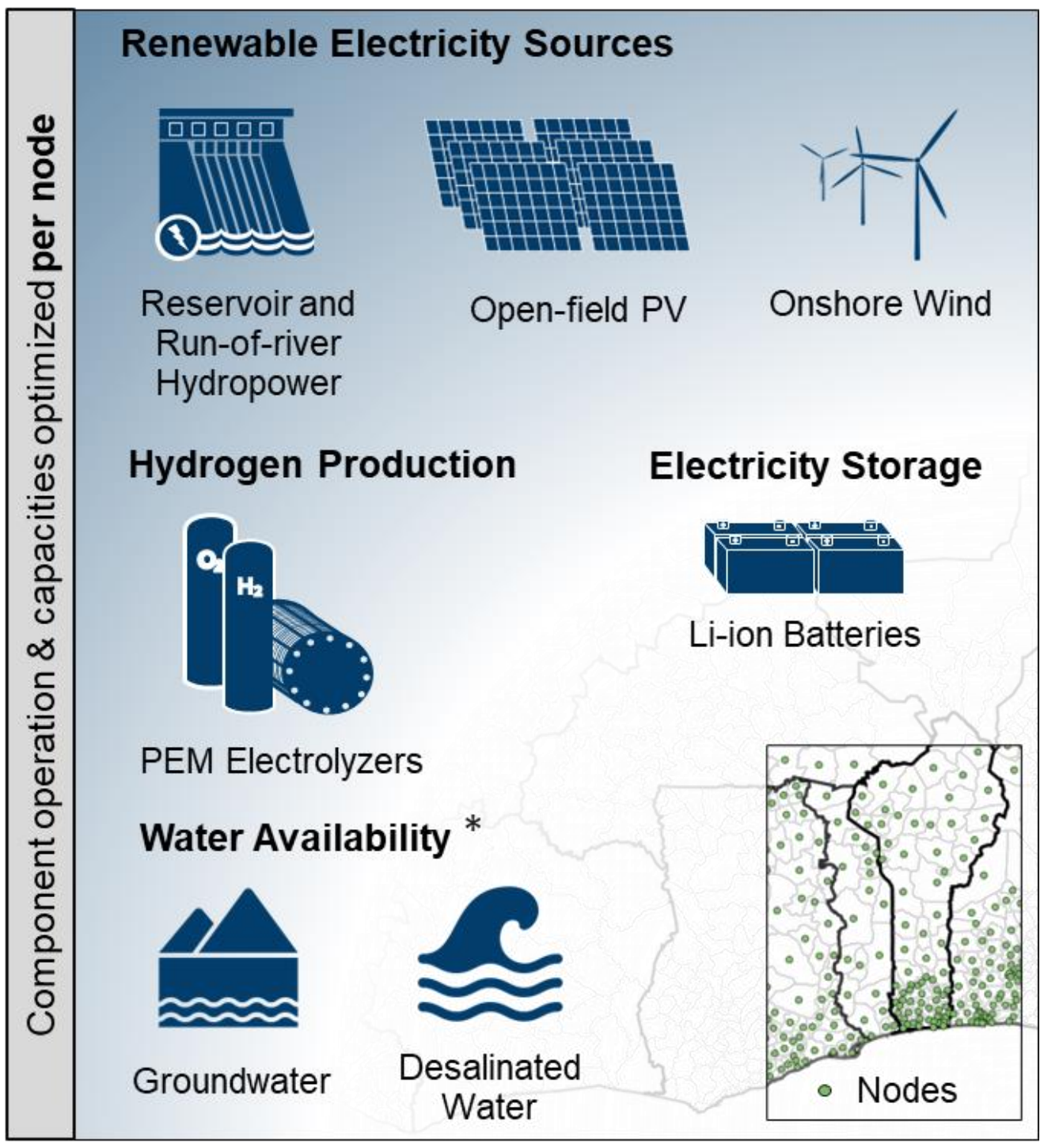

**Figure 2: Set of components added to each node of the model. *** *Icons made by Ayub Irawan and orvipixel from www.flaticon.com*

To determine the maximum amount of hydrogen that may be produced in each GID_2 region and to calculate the related levelized cost of hydrogen (LCOH), an independent energy system model was created for each region and optimized using the ETHOS.FINE (Framework for Integrated Energy System Assessment) optimization framework within the ETHOS model suite. The optimization framework was previously described by Welder et al. (Welder et al., 2018) and Groß et al. (Groß et al., 2024) and a publicly accessible version of the framework can be obtained from Github (FINE, 2024).

In the developed ETHOS.FINE model, each region is represented as one node, and an energy system is optimized for each node with several components. Onshore wind turbines, open-field PV parks, and existing hydropower plants were utilized as electricity sources, PEM electrolysis as green hydrogen production technology, and Li-ion batteries for electricity storage. Their assumed techno-economic parameters required for the optimization are listed in Table 2. For all these components the optimal capacity and operation in hourly resolution is optimized, while fulfilling an exogenously assumed gaseous hydrogen demand. The exogenous demand for hydrogen at each node incrementally raised in increments of 6% on average until the node's maximum hydrogen production capacity was depleted. The linear optimization minimized the annual cost of the necessary hydrogen production system and determining the most cost-effective technology portfolio for each node.

The renewable energy potential for onshore wind, open-field PV and the existing and planned hydropower capacities were aggregated from individual plant-level time-series to LCOE clusters per each GID_2 region. The LCOE range for wind turbines and PV within each GID_2 region was therefore split into 10 evenly spaced bins for each technology and the capacity-weighted average time series was calculated for each bin. In case of wind, the number of regional LCOE clusters was doubled to account for the high variation of wind potentials within even small geographical extent. As hydropower relies on few locations with possibly very different characteristics, especially when not depending on the same catchment area, the hydropower locations were not further clustered.

**Table 2. Techno-economic assumption parameters for each component (Stolten et al., 2021) (International Renewable Energy Agency, 2020)**

| Parameter | Investment [€/kW] | | | | Fix O&M [% of CAPEX] | Variable O&M [€/kWh] | Economic Lifetime |
|---|---|---|---|---|---|---|---|
| Year | 2020 | 2030 | 2040 | 2050 | 2020 - 2050 | 2020 - 2050 | 2020 - 2050 |
| Onshore Wind* | 1290 | 1130 | 1050 | 1000 | 2.5 | - | 20 |
| Open-field PV** | 690 | 450 | 370 | 320 | 1.7 | - | 20 |
| Run-of-river Hydropower | 1000 | 1000 | 1000 | 1000 | 2.5 | 0.005 | 40 |
| Reservoir Hydropower | 1700 | 1700 | 1700 | 1700 | 2.5 | 0.005 | 40 |
| PEM Electrolyzer | 800 | 500 | 400 | 350 | 3 | - | 10 |
| Li-ion Batteries | 311 | 175 | 153 | 131 | 2.5 | - | 15 |

*Onshore wind turbine values correspond to a reference turbine (capacity: 4.2 MW, rotor diameter: 136 m, hub height 120 m, avg. wind speed: 6.7 m/s) forming the basis for the optimized turbine designs

The projected local electricity and hydrogen demand should be prioritized over export countries to ensure local living standards and development as well as the acceptance of potential export schemes beyond local demand (Brauner et al., 2023). Therefore, local electricity and hydrogen demand is deducted from national hydrogen potentials in a subsequent step.

### 2.5 Socio-economic impact assessment

This study's research approach was developed using the OECD checklist for the reflective construction of composite indicators (OECD et al., 2008) and its main components are summarized in Table 3. Data sets were selected with a focus on sustainable development, identifying the sustainable development goals (SDGs) most influenced by green hydrogen projects, either directly or indirectly.

Firstly, the more direct impact would be on Goal 7 (affordable and clean energy) and Goal 8 (decent work and economic growth), as an increase in hydrogen projects leads to a concurrent increase in renewable energy capacity. Secondly, green hydrogen projects could indirectly affect goals 1 (no poverty), 2 (zero hunger), 3 (good health and well-being), and 13 (climate action). Economic growth and employment, ideally, could contribute to poverty reduction, enabling more individuals to afford food and healthcare. Additionally, the increase in green energy and clean fuel would have a positive impact on areas affected by indoor pollution and hazardous fossil fuel emissions, aligning with the imperative for climate action.

Further research, complemented by spatial data analysis, was conducted to assess the feasibility of socio-economic indicators in measuring the defined SDGs. The study also incorporated the prioritization of social development goals based on local visions, as revealed in surveys conducted during the project (Brauner et al., 2023).

**Table 3. Employed socio-economic composite indicators.**

| Sub-category | Indicator | Components | Unit |
|---|---|---|---|
| **Access to Energy** | Access to electricity | Population without access to electricity per GID 2 region's area | Capita/km² |
| **Macroeconomic effects** | Direct employment factor | Labour employment potential per installed capacity and per GID 2 region's area | Jobs/ (Mwp*km²) |
| **Other indirect effects** | Dependence on traditional biomass | Population using traditional biomass per GID 2 region's area | Capita/km² |
| | Population living under the poverty line | Share of the population under the poverty line | % |

To mitigate concerns about missing or insufficient data during both data selection and subsequent analysis, a robust approach was adopted, involving the collection and cross-checking of data from diverse sources. The selection process encompassed compiling single data points and historical records within the African context to ensure the most reliable data for the chosen indicators.

## Mapping Energy Access

The indicator for access to energy began with the recognition that data on energy access is typically available only at the national level for both urban and rural areas (Ritchie et al., 2019). By leveraging population density data, urbanization could be defined. To ascertain the regional distribution of the population without access to energy, a correlation between density and access was established for each ADM-2 region and validated using data from (Falchetta et al., 2020):

$$\mathrm{I}_1 = \left(\mathrm{EAu}^{\mathrm{ADM-1}} * \sum_{\mathrm{ADM-2}} \mathrm{PoPu} + \mathrm{EAr}^{\mathrm{ADM-1}} * \sum_{\mathrm{ADM-2}} \mathrm{PoPr}\right) / \mathrm{Area}$$

This distribution was calculated using the urban energy access rate at the ADM-1 level ($\mathrm{EAu}^{\mathrm{ADM-1}}$), the rural energy access rate at the ADM-1 levels ($\mathrm{EAr}^{\mathrm{ADM-1}}$), combined with the total urban and rural population per area at the ADM-2 level ($\mathrm{PoPu}$ *and* $\mathrm{PoPr}$ *respectively*).

## Mapping Local Employment

Employment figures were calculated using regional multipliers for various renewable energy and power-to-hydrogen technologies. The unemployment rate multiplied by the regional labor force available in ADM-2 regions was used to calculate employment for the area, as described by the following equation:

$$\mathrm{I}_2 = \mathrm{RM} * \left({}^{\text{average}}_{\text{RES}}\,\mathrm{EF} + \mathrm{EF}_{\mathrm{PtH}}\right) * \mathrm{UP} * \mathrm{Labor/Area}$$

Here, RM is the regional multiplier, ${}^{\text{average}}_{\text{RES}}\,\mathrm{EF}$ represents the global average employment factor for various renewable energy sources, including wind, solar, and hydro, and $\mathrm{EF}_{\mathrm{PtH}}$ representing the global employment factor for power-to-hydrogen technologies. The product is then multiplied by $\mathrm{UP}$ (unemployment rate) and Labor (total labor force, i.e., the population between the ages of 15 and 64).

### Other Impacts

The methodology for calculating the population using traditional biomass mirrored that used for mapping energy access but utilized clean fuel access numbers instead (Ritchie et al., 2019). Regarding the poverty indicator, the share of the population under the poverty line for ADM-2 regions was considered, like ADM-1 regions (Hai-Anh H. Dang et al., 2022).

### Normalization and Weighting

Various normalization techniques, as per OECD guidelines (Union & Centre, 2008), were applied to the set of indicators. Standardization (or z-scores) was found to best reflect the indicator scores. Equal weighting was applied to direct effects, with less importance given to the indirect ones, justified by the absence of other known means of weighting (Ebert & Welsch, 2004). Arithmetic means were used for aggregating various composite indicators and sub-indexes.

## 2.6 Result dissemination via web-based geographical user interface

To ensure widespread accessibility of the study's findings, we developed a user-friendly GUI as a web application (Jülich Systems Analysis, 2024). The GUI showcases the obtained results, presenting them on a map through color-coded layers where visualization of various factors influencing the potential of green hydrogen in select regions across Africa is facilitated. Users can also engage with the map interface to delve into the data, enabling them to glean insights that foster comprehension of socio, technical and economic aspects of it.

The GUI architecture leverages Docker-based containerization (Docker, 2024), which facilitates a modular and scalable application structure. The frontend employs React (React, 2024) and utilizes the Mapbox API (Mapbox, 2024) to provide dynamic mapping functionalities, allowing users to explore complex geographical data intuitively. The backend integrates Node.js (Node.js, 2024) for server-side operations, paired with a PostgreSQL database enhanced with PostGIS (PostGIS, 2024). This setup effectively manages and queries spatial data, supporting advanced spatial queries essential for the study. A reverse proxy via Nginx (Nginx, 2024) optimizes deployment and enhances performance by managing traffic efficiently. To accurately handle and display geographic data, we employ various software solutions including Geokit (GeoKit, 2024), GeoPandas (Joris Van den Bossche et al., 2024), and QGIS (QGIS.org, 2024). These tools guarantee precise geographic positioning of the data through established standards like EPSG:3857, also referred to as Web Mercator (European Petroleum Survey Group Geodesy, 2020), and EPSG:4326, commonly known as WGS84 (European Petroleum Survey Group Geodesy, 2022).

This framework guarantees a functional and user-friendly GUI that maintains precise and informative visualizations. It enables effective communication of complex data, enhancing understanding of green hydrogen potential across regions. Through this advanced GUI, stakeholders access crucial data, aiding decisions on sustainable energy development in Sub-Saharan Africa.

## 3 Case Study for Selected Regions

For showing which results can be obtained with the developed multidisciplinary approach one exemplary region for each type of result was selected. The “Ouémé” region in Benin was chosen to illustrate an exemplary land eligibility and renewable energy simulation approach as well as the water availability and socio-economic assessment. The hydrogen modeling approach is then explained based on an exemplary hypothetical region in order to demonstrate all aspects of interest such as groundwater extraction, seawater desalination, utilization of all types of renewable sources of electricity etc. within a single region.

### 3.1 Land eligibility assessment for open-field photovoltaic and onshore wind turbines

Figure 3 displays the distribution of local preferences for buffer values of the land eligibility assessment, specifically for the "Leisure and Camping" criterion, which is one of the 33 criteria mentioned in Table 1. Respondents from twenty-two out of the thirty-one surveyed countries in Sub-Saharan Africa submitted their preferences, represented by dots on the graph. For onshore wind, most countries chose a buffer value that is close to the median, although some countries preferred a higher buffer value, and a few countries chose a lower buffer value. On the other hand, for open-field PV, none of the countries chose values close to the median buffer value. Most countries selected a buffer value lower than the median, with only a few countries choosing a higher buffer value. These buffer values corresponding to each criterion are applied to all GID_2 regions within a country, unless otherwise stated, and hence also to all regions in Benin including the “Ouémé” region when the final eligibility for the region is be computed.

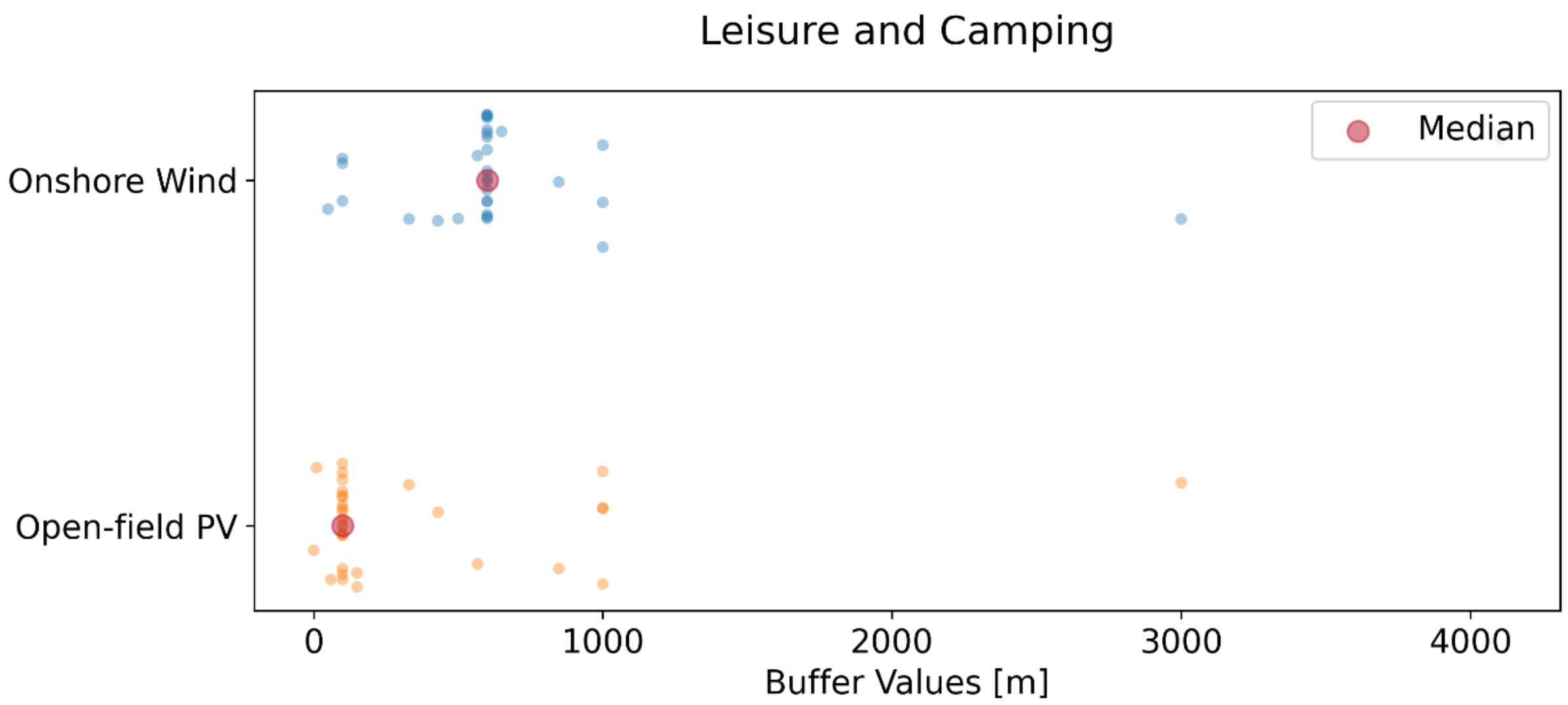

**Figure 3. Distribution of buffer values received for the "Leisure and Camping" criterion from the surveyed countries in Sub-Saharan Africa**

### 3.2 Renewable energy potential assessment

Wind turbines and PV modules are placed within the identified exemplary eligible area (based only on three criteria “roads”, “settlements” and “forests”) which was shown in Figure 1. Figure 4 displays the resulting distribution of onshore wind placements and large-scale open-field PV parks. This positioning method has resulted in about 696 eligible locations for the placement of onshore wind turbines for the exemplary region Ouémé (BEN.10_1) in Benin. The total available area for open-field PV placement, after the distribution of polygons, is estimated to be around 318 km$^2$. This translates to a total capacity of 2.45 GW of onshore wind potential and to approximately 16 GW of open-field PV potential for this specific region.

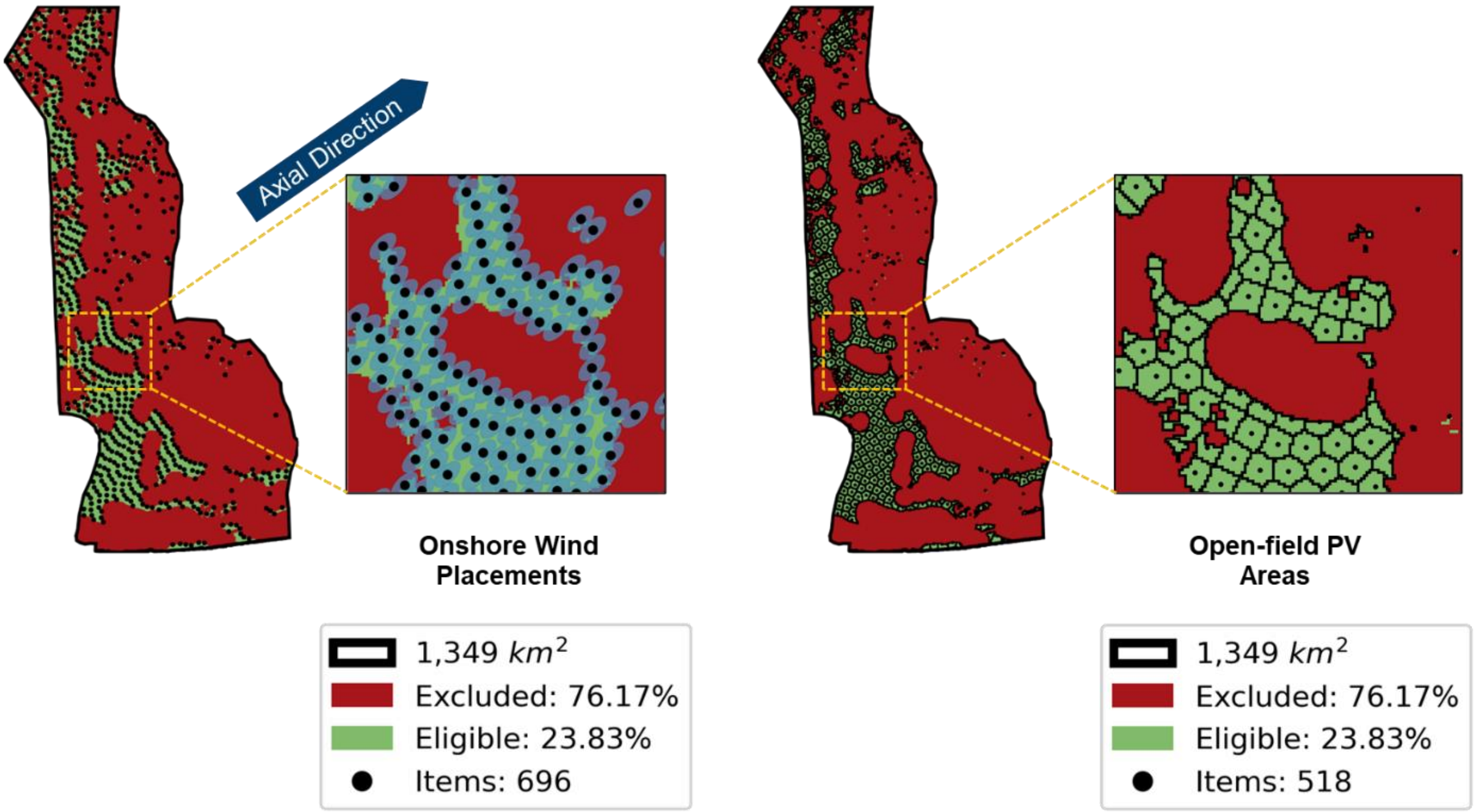

**Figure 4. Distribution of onshore wind turbines (left) and PV area (right) placements within the eligible areas**

Figure 5 displays the distribution of levelized cost of electricity for onshore wind turbines and open-field PV parks across the locations shown in Figure 4. In the featured region, the cost distribution of electricity generated by onshore wind turbines is relatively steep, ranging from approximately 14 Ct$_€$/kWh to 31 Ct$_€$/kWh. Conversely, electricity produced by distributed open-field PV parks is comparatively inexpensive and has a lower cost spread ranging from ~2.9 Ct$_€$/kWh to 3.1 Ct$_€$/kWh in 2030.

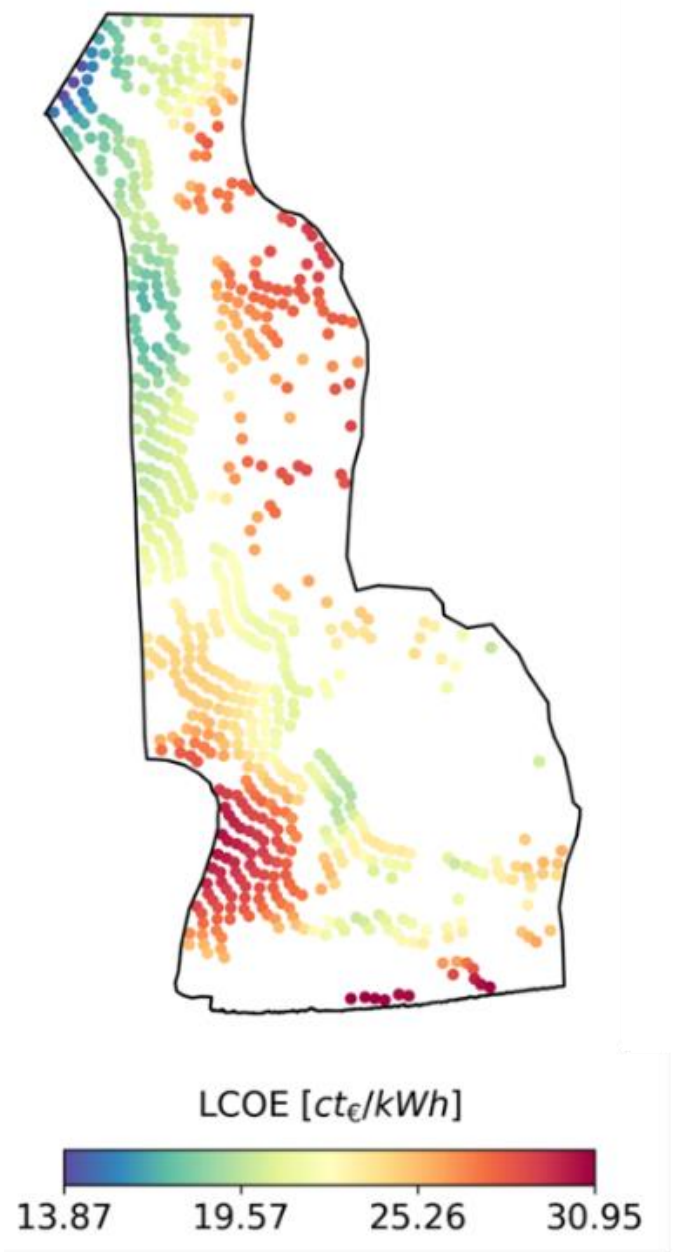

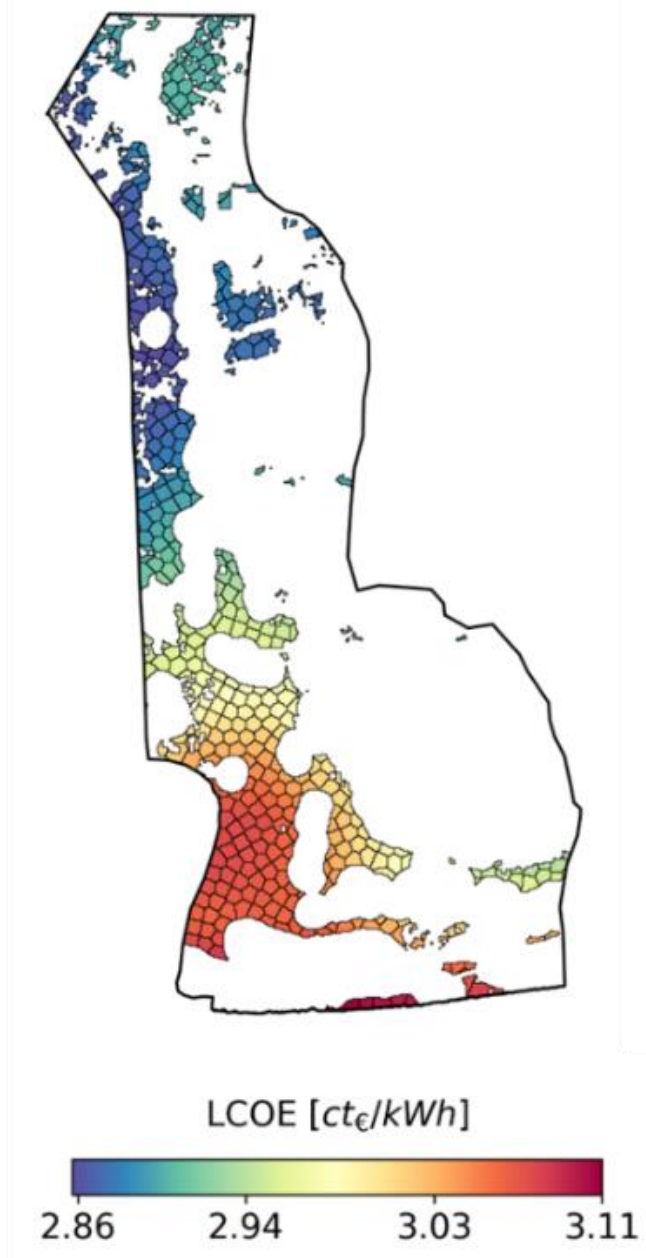

**Figure 5: Levelized cost of electricity (LCOE) for the exemplary onshore wind placements (left) and open-field PV placements (right) corresponding to assumptions for the year 2030**

### 3.3 Sustainable water supply assessment

### 2020 groundwater sustainable yield

Reflecting groundwater sustainable yield in 2020, Figure 6 illustrates long-term average (2015 – 2035) groundwater sustainable yield maps for the selected region. These maps present two climate scenarios: RCP2.6 (Figure 6a & c & e) and RCP8.5 (Figure 6b & d & f). Within each scenario, three cases are investigated: conservative (Figure 6 a & b), medium (Figure 6 c & d), and extreme conditions (Figure 6e & f). Figure 6 reveals a noteworthy trend: the availability of water increases progressively from conservative to extreme scenario. As Figure 6 shows, the average estimates of 2020 groundwater sustainable yield in the selected region for the year 2020 under RCP2.6 (RCP8.5) scenarios would be 5.5 (3.4) mm $yr^{-1}$ (in the conservative scenario), 63.3 (49.6) mm $yr^{-1}$ (in the medium scenario), and 150 (120.6) mm $yr^{-1}$ (in the extreme scenario). The simulation for the medium case (Figure 6c & d) positioned between the lower (conservative case) and upper (extreme case) boundaries of available groundwater resources is an ideal and optimum option for hydrogen production.

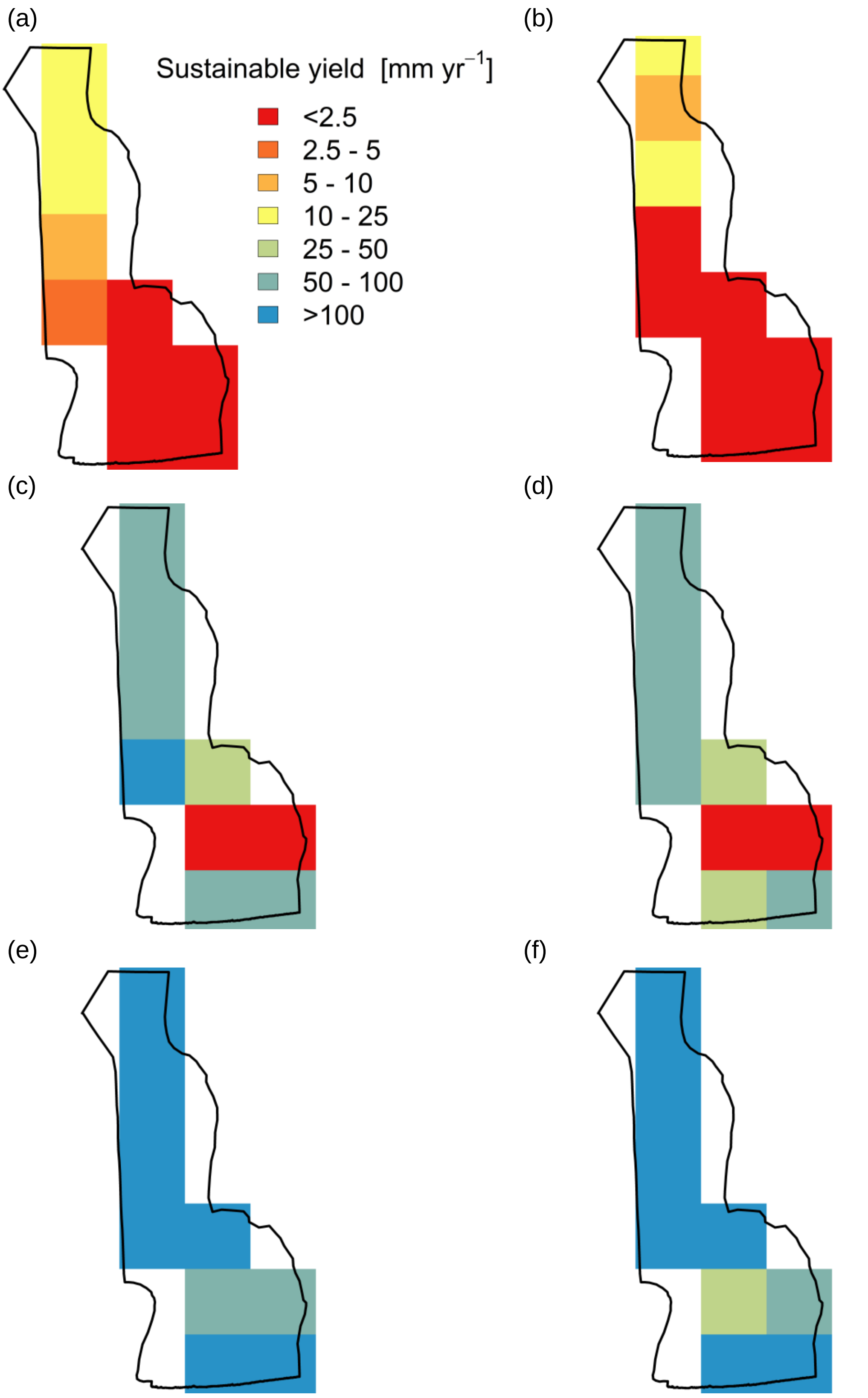

**Figure 6:** Groundwater sustainable yield representative for the year 2020 calculated as the long-term average (2015 - 2035) of simulations, taking into account two climate scenarios: RCP2.6 (a & c & e) and RCP8.5 (b & d & f). Within each scenario, three cases are investigated: conservative (a & b), medium (c & d), and extreme conditions (e & f).

Table 4: The average estimates of groundwater sustainable yield in the selected region for 2020 (2015 - 2035), 2030 (2015 - 2045) and 2050 (2036 - 2065) considering two climate scenarios: RCP2.6 and RCP8.5 under conservative, medium, and extreme conditions.

| Scenario | Groundwater sustainable yield [mm $yr^{-1}$] | | | | | |
|---|---|---|---|---|---|---|
| | 2020 | | 2030 | | 2050 | |
| | RCP2.6 | RCP8.5 | RCP2.6 | RCP8.5 | RCP2.6 | RCP8.5 |
| Conservative | 5.5 | 3.4 | 3.1 | 2.1 | 0.2 | 0 |
| Medium | 63.3 | 49.6 | 49.3 | 38.3 | 29.6 | 19.5 |
| Extreme | 150 | 120.6 | 122.2 | 100.9 | 82.8 | 54 |

**2030 groundwater sustainable yield**

Long-term average (2015 - 2045) groundwater sustainable yield maps representative for the year 2030 are shown in Figure 7, taking into account two climate scenarios: RCP2.6 (Figure 7a & c & e) and RCP8.5 (Figure 7b & d & f) and three cases: conservative (Figure 7a & b), medium (Figure 7c & d), and extreme conditions (Figure 7e & f). The regional assessment (Figure 7) revealed that across the entire area, the average groundwater sustainable yield for the year 2030 under RCP2.6 (RCP8.5) scenarios would be 3.1 (2.1) mm $yr^{-1}$ (in the conservative scenario), 49.3 (38.3) mm $yr^{-1}$ (in the medium scenario), and 122.2 (100.9) mm $yr^{-1}$ (in the extreme scenario).

**2050 groundwater sustainable yield**

Nonetheless, long-term average (2036 - 2065) groundwater availability representative for 2050 (Figure 8) under both RCP2.6 (Figure 8a & c & e) and RCP8.5 (Figure 8b & d & f) scenarios experienced a notable decrease across all three cases: conservative (Figure 8a & b), medium (Figure 8c & d), and extreme conditions (Figure 8e & f), in comparison to 2020 (Figure 6) and 2030 (Figure 7). On average, as indicated in Figure 8, the entire region is projected to demonstrate 2050 mean groundwater sustainable yield values under RCP2.6 (RCP8.5) scenarios equal to 0.2 (0) mm $yr^{-1}$ (in the conservative scenario), 29.6 (19.5) mm $yr^{-1}$ (in the medium scenario), and 82.8 (54) mm $yr^{-1}$ (in the extreme scenario) which are significantly lower than those observed in 2020 and 2030.

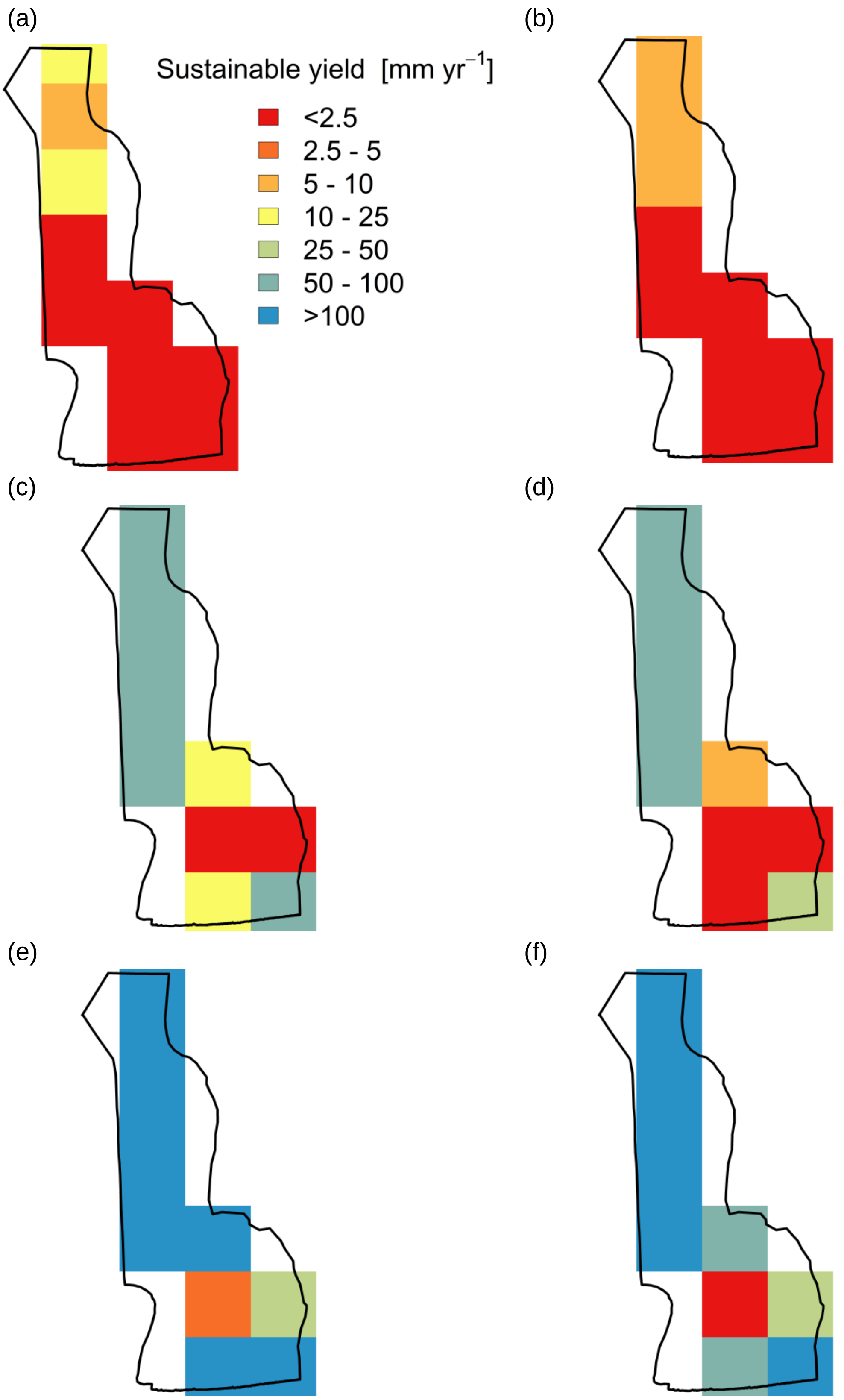

**Figure 7.** Groundwater sustainable yield representative for the year 2030 calculated as the long-term average (2015 - 2045) of simulations, taking into account two climate scenarios: RCP2.6 (a & c & e) and RCP8.5 (b & d & f). Within each scenario, three cases are investigated: conservative (a & b), medium (c & d), and extreme conditions (e & f).

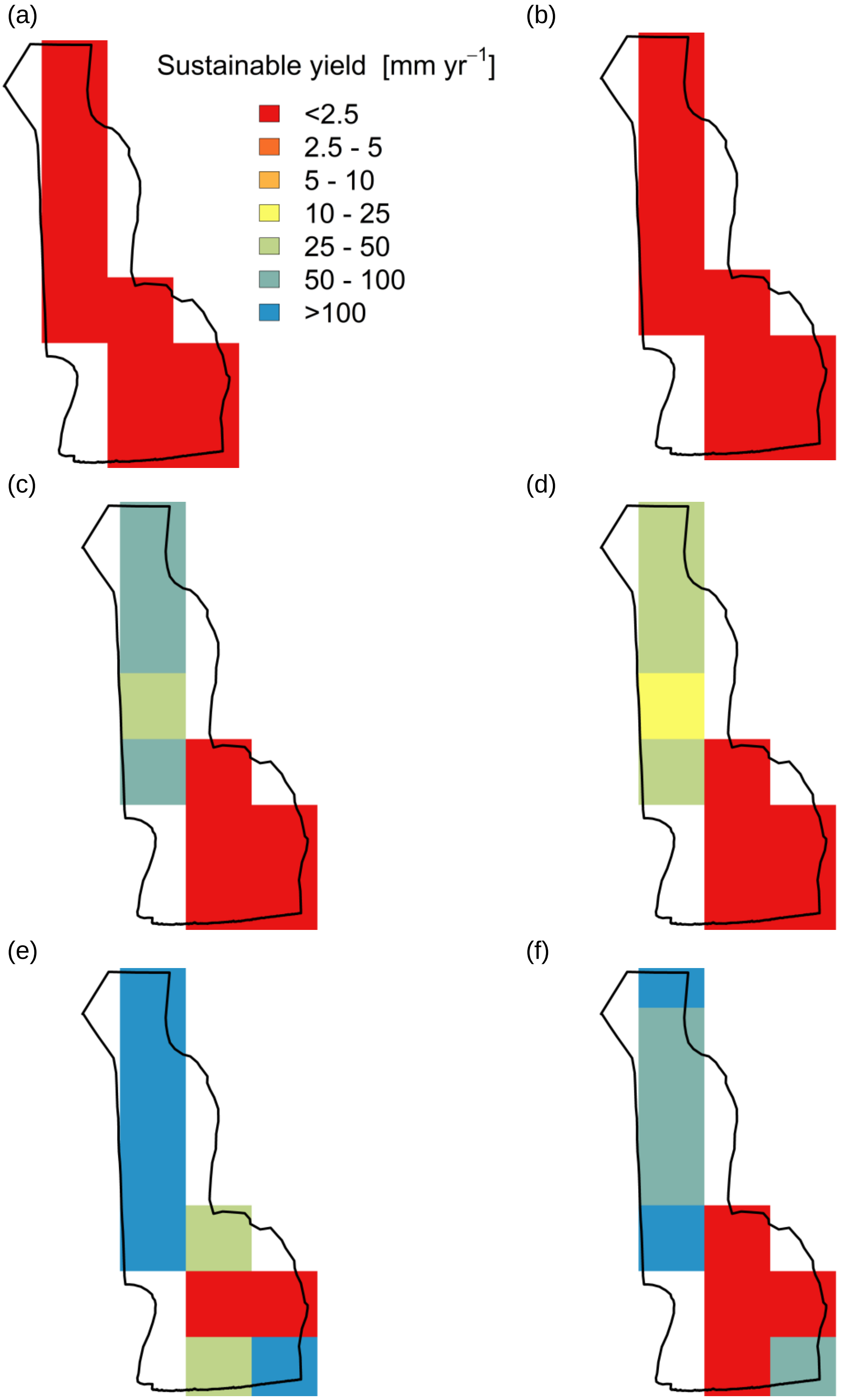

**Figure 8:** Groundwater sustainable yield representative for the year 2050 calculated as the long-term average (2035-2065) of simulations, taking into account two climate scenarios: RCP2.6 (left panels) and RCP8.5 (right panels). Within each scenario, three cases are investigated: conservative (a & b), medium (c & d), and extreme conditions (e & f).

### 3.4 Local green hydrogen potential assessment

The levelized cost of hydrogen (LCOH) and the respective hydrogen production potential for each analyzed expansion step in an example region in 2050 is shown in Figure 9. For this step, a hypothetical region was defined that demonstrates all aspects of interest, in particular realistic shares of potentials for all selected technologies except geothermal power and an existing yet limited sustainable groundwater availability. The region has a maximum technical hydrogen potential of 2391 $kt_{H2}/a$, with 50% of the hydrogen being producible from local groundwater (see grey line in Figure 9). The optimized energy system composition resulting from the cost contribution of the different technologies involved is displayed in the form of a bar at each expansion step. At lower levels of exploitation of the maximum potential, up to about 2%, the electricity required for the hydrogen production is supplied by hydropower (reservoir). Starting from the 4% expansion step and up until the 90% expansion step, in addition to hydropower gradually more and more open-field PV capacity is utilized. Starting from the 96% expansion step, gradually increasing amounts of onshore wind capacity as well as batteries are added to the mix until the 100% expansion step is reached. This shows that onshore wind is the most expensive option to produce hydrogen in this region and additionally requires battery storage.

The LCOH starts at about 1.3 €/$kg_{H2}$ for pure hydropower production and lies at that value up until the 4% expansion step. Including open-field PV into the system increases the LCOH to roughly 2.1 €/$kg_{H2}$, while utilizing onshore wind drives the LCOH gradually to about 2.4 €/$kg_{H2}$ at the 100% expansion step. Below 50% of the maximum potential, local groundwater is used for hydrogen production. Above that, desalination must be used to provide freshwater for the electrolysis. The cost of water supply is negligible though, representing less than 1.2% of the total levelized cost of hydrogen despite the partial use of desalination.

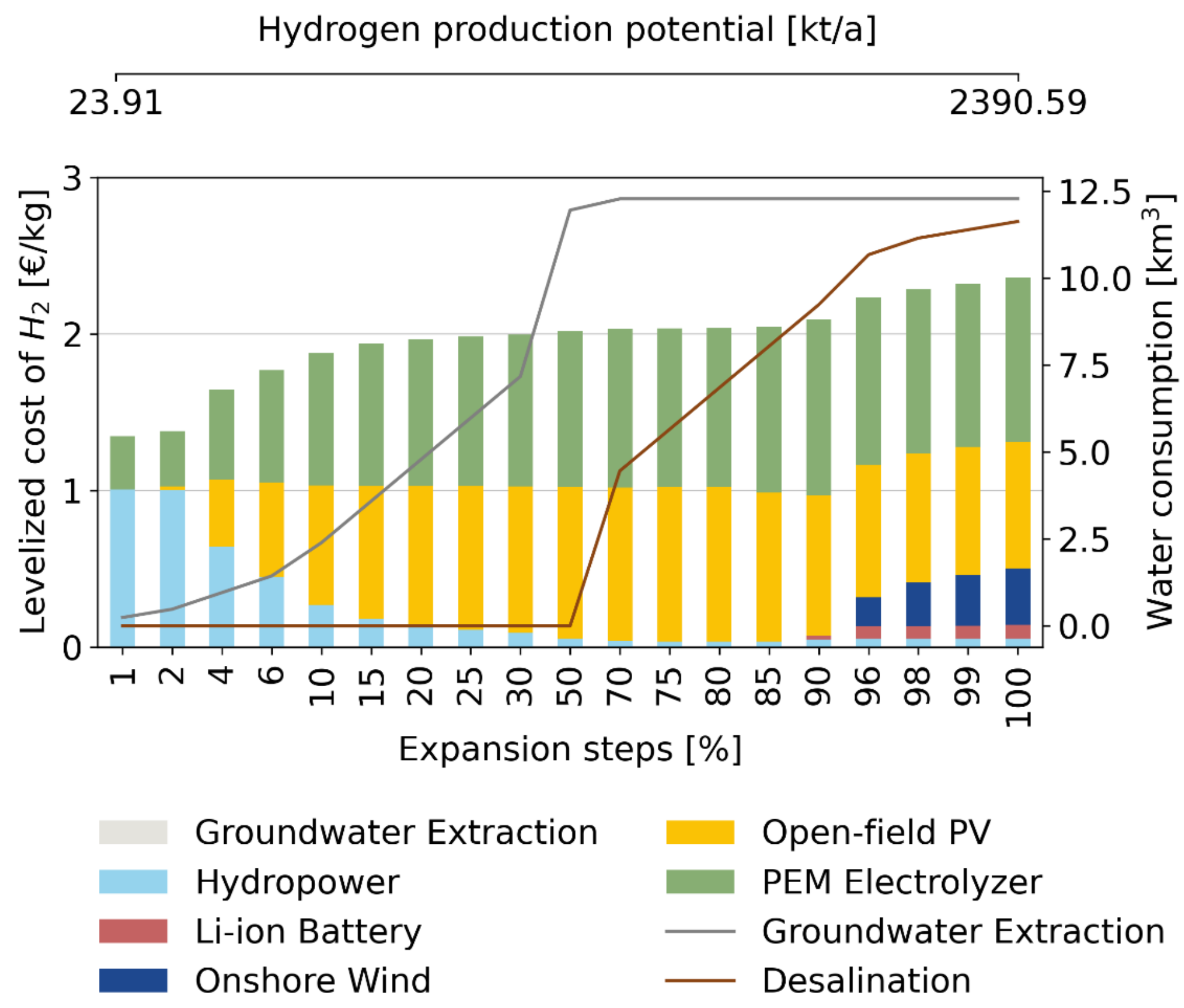

**Figure 9. Cost shares of each technology at each expansion step within the hypothetical region for 2050.**

### 3.5 Socio-economic impact assessment

Here again, the example of Ouémé in Benin is chosen to illustrate the methodology used for the socio-economic impact assessment as shown in Figure 10. It shows how the results are synthesized at ADM-2 regions resolution to assess a socio-economic indicator evaluating the local impact of projects, ranging from very low to very high. Initially, population density at a 1 km resolution is employed to differentiate between urban and rural densities through the urbanization rate. This differentiation is used to determine the population lacking access to energy, including electricity and clean fuel, at a 1 km resolution, utilizing rural and urban access rates at the regional level, and national level if regional data is unavailable. Subsequently, these findings are aggregated at the ADM-2 grid to establish the population density without access to electricity per administrative grid area.

The macroeconomic impact is derived from labor density, which is computed from the same urban and rural population densities, adjusted by the average available labor force and unemployment rate at the regional or national level. For Ouémé, the labor force percentage fluctuates between 52% and 58%, with a corresponding regional unemployment rate of 1.6%. This is then multiplied by the regional average employment factor for various technologies. For instance, in the ADM-1 region of Figure 10, this translates to 5.1 jobs/MWp for PV, 3.2 jobs/MWp for onshore wind, 5.9 jobs/MWp for hydro, and 1.7 jobs/MWp for P2H. Additionally, the percentage of population below the poverty line in the region corresponds to an aggregated average of 52 %. This is incorporated in addition to the macroeconomic effects and energy access indicators using a weighted average approach.

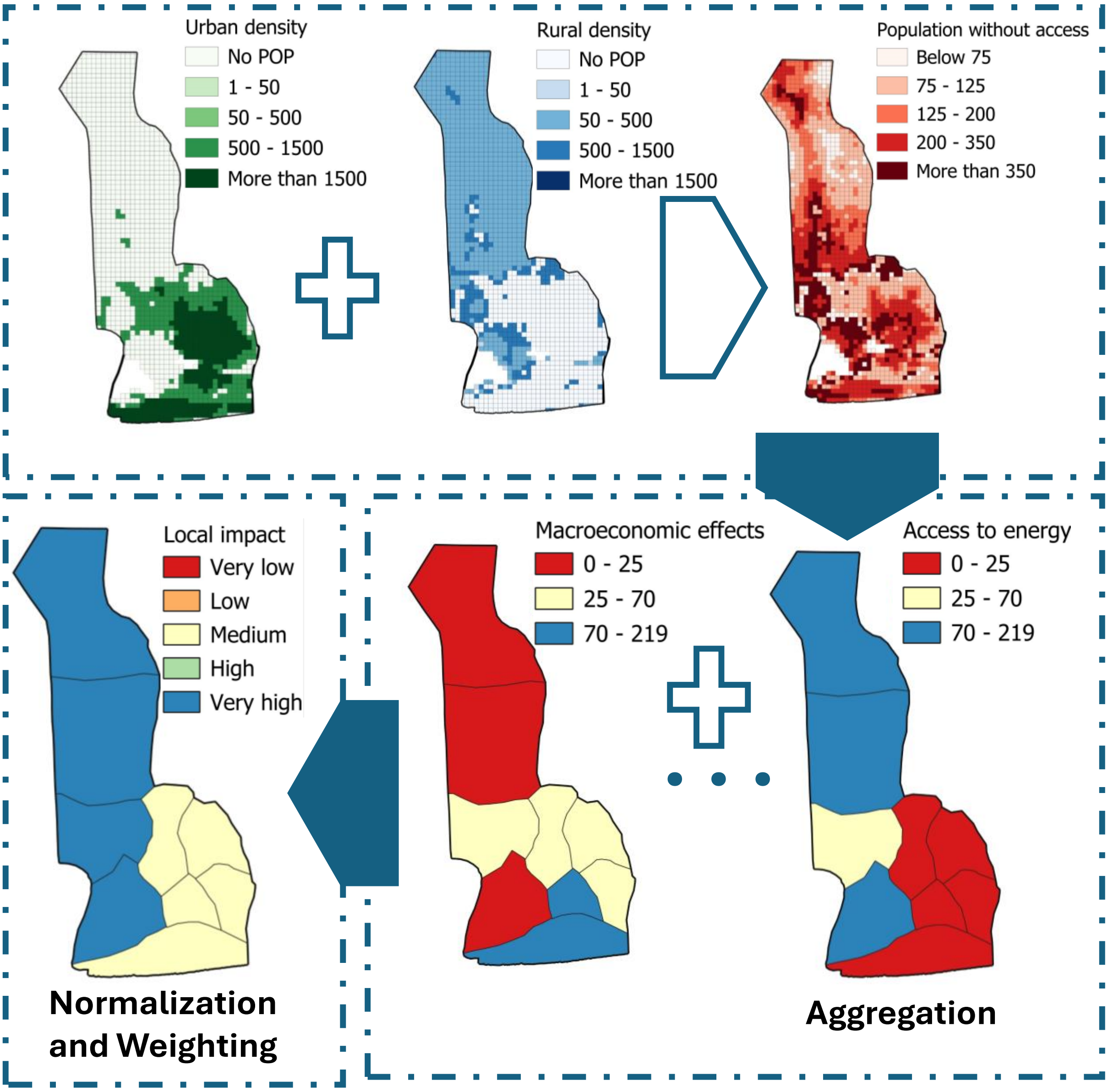

**Figure 10 Example of Ouémé for the construction of the socio-economic local impact indicator**

## 4 Conclusions

By utilizing a multidisciplinary approach to cover all crucial dimensions of green hydrogen in Sub-Saharan Africa data, a robust decision support can be provided. This stems not only from the combination of tools but from the advances contained in each step as well. This ranges from the first systematic inclusion of local preferences for land eligibility for renewable energy technologies in Sub-Saharan Africa combined with the best available land cover data and the high level of detail in the renewable energy potential assessment with specific placements, to the achieved advances in calculating sustainable groundwater yields under different climate scenarios together with considering seawater desalination with water transport for regions without or not sufficient sustainable groundwater yield. Finally, those advances feed-in the new approach to derive the green hydrogen potentials locally with highly resolved optimizing energy system models, which consider the variable nature of the involved renewable energy technologies and their interplay with the other components within the energy system, especially the sizing of electrolysis and batteries. This techno-economic approach under environmental constraints is complemented with a social perspective to enrich the decision support. Each of those steps bear unique advances and options for future applications and improvements.

For the comprehensive land eligibility approach, we used 33 criteria with a high spatial resolution to determine eligible land areas for onshore wind and open field PV. In coordination with local partner in these regions, we determined their local preferences for exclusions. We found that local preferences of land eligibilities and their buffer distances can vary between -100% and +1000% based to averaged values. Therefore, it is important to consider the local sociopolitical preferences to be able to conduct accurate land eligibility analysis.

The approach for the open field PV and onshore wind power for countrywide potential assessment utilizes location specific placements and simulations of PV parks and wind turbines. It can be found that the cost of renewable energies and their range varies between 14 $Ct_{€}$/kWh to 31 $Ct_{€}$/kWh for wind power and ~2.9 $Ct_{€}$/kWh to ~3.1 $Ct_{€}$/kWh for PV within the exemplary region in Benin. It is also shown that the potentials vary largely within each region. Therefore, a detailed location specific approach as described in this work is necessary to determine the renewable energy potential in the context of hydrogen potential analysis.

Groundwater availability and cost are considered quantitatively at regional scale for the first time in an energy system model and the maximum sustainable shares of groundwater and additionally required freshwater from desalination are calculated. Utilizing land surface modeling stands as a viable avenue to assess water balance elements, compute groundwater recharge, and generate maps depicting groundwater sustainable yield. This sustainable yield essentially denotes the remaining water volume after accounting for all existing human requirements and potential environmental water needs. Therefore, these quantitative representations serve as valuable information for evaluating the sustainability of African groundwater resources, offering insights not only into current conditions but also potential scenarios under future climate change projections. As advancements in hydrological modeling continue to evolve and interdisciplinary approaches gain traction, future groundwater analysis may increasingly rely on integrated sub-surface-surface-atmosphere models incorporating real-time data streams from in situ and satellite observations and utilizing machine learning algorithms to enhance predictive capabilities and improve the sustainable management of groundwater resources.

The inclusion of water limitations in the energy system approach shows, that cost of water supply proves negligible in the overall Levelized Cost of Hydrogen (LCOH), which ranges from roughly 1.3 €/$kg_{H2}$ to 2.4 €/$kg_{H2}$ in 2050. Still, careful consideration must be given to the selection of water sources, with options ranging from seawater to groundwater. It should be highlighted that this analysis assumes dry cooling based on Holst et al. (Holst et al., 2021). Depending on the selected cooling process, additional water demand for electrolysis may incur, then leading to a total water demand of 17.5 to 95 $l_{H20}$/$kg_{H2}$ (Global Alliance Powerfuels, 2021; IRENA & Bluerisk, 2023; Kabir et al., 2023). Moreover, this study focuses on local energy systems for hydrogen production. Therefore, synergies between neighboring regions are neglected compared to Franzmann et al. (Franzmann et al., 2023).

Regarding the socio-economic dimension of green hydrogen in Sub-Saharan Africa the utilization of the OECD checklist for constructing composite indicators ensured a robust research approach, providing clear direction by focusing on sustainable development goals (SDGs) influenced by green hydrogen projects. This is done based on the analysis of direct and indirect impacts of green hydrogen projects on various SDGs, coupled with spatial data analysis, which enhanced the understanding of socio-economic indicators' feasibility and relevance. Moreover, the integration of local visions through surveys added depth to the Sub-Saharan African vision regarding the socio-economic indicators that should be prioritized. This is ensured by cross-checking from diverse and local sources as well to complete the data collection and mitigate concerns about missing or insufficient data, ensuring reliability. Through this approach the local impact of green hydrogen and renewable energy sources projects are considered for the first time at the regional scale in the Sub-Saharan African case. This is based on macroeconomic effects measured by the direct local employment of potential projects and the indirect promotion of access to electricity and energy via the direct use of renewable energy sources. The main driver of local impact lies in promoting energy access and employing regional labor for construction and operations. However, long-term considerations including the growth of regional renewable industries, necessitates future data improvements to account for broader job and local impacts, such as those induced from infrastructure, manufacturing and export projects associated to a green hydrogen economy.

Finally, the results achieved for the case study regions not only prove the applicability of the newly developed approach presented in this paper but the benefits arising from the multidisciplinary approach, too. Those stretch from the value of local preferences for placing renewable energies across endogenous consideration of sustainable water supply to finally identifying main drivers of local socio-economic impacts of green hydrogen. Now a large-scale application will be the next logical step.

## Acknowledgements

A major part of this work has been carried out within the framework of the $H_2$ Atlas-Africa project (03EW0001) funded by the German Federal Ministry of Education and Research (BMBF). Additionally, we acknowledge funding by the European Space Agency (ESA) in the Framework of the Dragon 5 cooperation between ESA and Chinese Ministry of Science and Technology under Projects 59197 and 59316. Special thanks to the partners in the focus countries led by the national team leaders for fruitful discussions and providing local data: Mr. Chipilica Barbosa (Angola), Prof. Julien Adounkpe (Benin), Dr. Lapologang Magole (Botswana), Prof. Tanga Pierre Zoungrana (Burkina Faso), Prof. Luis Jorge Fernandes (Cape Verde), Mr. Simphiwe Khumalo (Eswaitni), Prof. Wilson A. Agyare (Ghana), Prof. Konate Souleymane (Ivory Coast), Mr. Joseph Kalowekamo (Malawi), Dr. Yacouba Diallo (Mali), Mr. Mohamed Abdoullah Muhamadou (Mauritania), Dr. Pradeep M. K. Soonarane (Mauritius), Prof. Boaventura Chongo Cuamba (Mozambique), Mr. Panduleni Hamukwaya (Namibia), Prof. Rabani Adamou (Niger), Prof. Apollonia Okhimamhe (Nigeria), Dr Aime Tsinda (Rwanda), Dr. Ibrahima Barry (Senegal), Mr. Crescent Mushwana (South Africa), Mr. Mathew Matimbwi (Tanzania), Prof. Sidat Yaffa (The Gambia), Prof. Agboka Komi (Togo), Mr. Edson Twinomujuni (Uganda), Dr. Martin Mbewe (Zambia), and Dr. Fortunate Farirai (Zimbabwe). We thank also Mrs. Alberta Aryee, Mrs. Chenai Marangwanda, and Dr. Imasiku Katundu for their valuable support and facilitating interaction with the national teams in the various countries of West and Southern Africa.

## Author contributions

Conceptualization: HH; methodology – land eligibility: CW, SI, HH; methodology – renewable energy assessment: EUPS, CW, SI, DF, HH; methodology – sustainable groundwater yield: BB, BO, CM, HV, HJHF; methodology – seawater desalination and water transport: CW, HH;

methodology – local green hydrogen potential assessment: CW, HH; methodology – socio-economic impact assessment: AL, WK, SV; writing – original draft: HH, SI, BB, BO, CM, HV, HJHF, EUPS, DF, CW, AL, SA; writing – review and editing: all named authors; visualization: SI, BB, DF, CW, EUPS, AL; supervision: HH, JL, and DS. All authors have read and agreed to the published version of the manuscript.